\documentclass[aps,preprintnumbers,prl,twocolumn,superscriptaddress]{revtex4}

\usepackage{epsfig,latexsym,cancel,amssymb,amsmath}
\usepackage{graphicx}
\usepackage{epstopdf}
\usepackage{color}

\def\be{\begin{equation}}
\def\ee{\end{equation}}
\def\bea{\begin{eqnarray}}
\def\eea{\end{eqnarray}}

\def\nn{\nonumber}
\def\bbuildrel#1_#2^#3{\mathrel{\mathop{\kern 0pt#1}\limits_{#2}^{#3}}}
\def\slash#1{\setbox0=\hbox{$#1$}#1\hskip-\wd0\dimen0=5pt\advance
       \dimen0 by-\ht0\advance\dimen0 by\dp0\lower0.5\dimen0\hbox
         to\wd0{\hss\sl/\/\hss}}

\newcommand{\gae}{\lower 2pt \hbox{$\, \buildrel {\scriptstyle >}\over {\scriptstyle
\sim}\,$}}
\newcommand{\lae}{\lower 2pt \hbox{$\, \buildrel {\scriptstyle <}\over {\scriptstyle
\sim}\,$}}

\newcommand{\beq}{\begin{eqnarray}}
\newcommand{\eeq}{\end{eqnarray}}

\newcommand{\ba}{\begin{array}}
\newcommand{\ea}{\end{array}}

\newcommand{\CPv}{{\textrm{\fontsize{6}{11}\selectfont CP}\!\!\!\!\!\!\!\diagup}}

\long\def\symbolfootnote[#1]#2{\begingroup%
\def\thefootnote{\fnsymbol{footnote}}\footnote[#1]{#2}\endgroup}

\newcommand{\CPV}{{CP\!\!\!\!\!\!\!\!\raisebox{0pt}{\small$\diagup$}}}

\def\lsim{\mathrel{\rlap{\lower4pt\hbox{\hskip1pt$\sim$}}
    \raise1pt\hbox{$<$}}}         
\def\gsim{\mathrel{\rlap{\lower4pt\hbox{\hskip1pt$\sim$}}
    \raise1pt\hbox{$>$}}}         

\def\lsim{\:\raisebox{-0.5ex}{$\stackrel{\textstyle<}{\sim}$}\:}
\def\gsim{\:\raisebox{-0.5ex}{$\stackrel{\textstyle>}{\sim}$}\:}

\def\issue(#1,#2,#3){{\bf #1}, #2 (#3)}
\def\opcit(#1){ {\em op. cit.}, #1}

\def\APP(#1,#2,#3){Acta Phys.\ Polon.\ \issue(#1,#2,#3)}
\def\ARNPS(#1,#2,#3){Ann.\ Rev.\ Nucl.\ Part.\ Sci.\ \issue(#1,#2,#3)}
\def\CPC(#1,#2,#3){Comp.\ Phys.\ Comm.\ \issue(#1,#2,#3)}
\def\CIP(#1,#2,#3){Comput.\ Phys.\ \issue(#1,#2,#3)}
\def\EPJC(#1,#2,#3){Eur.\ Phys.\ J.\ C\ \issue(#1,#2,#3)}
\def\EPJD(#1,#2,#3){Eur.\ Phys.\ J. Direct\ C\ \issue(#1,#2,#3)}
\def\IEEETNS(#1,#2,#3){IEEE Trans.\ Nucl.\ Sci.\ \issue(#1,#2,#3)}
\def\IJMP(#1,#2,#3){Int.\ J.\ Mod.\ Phys. \issue(#1,#2,#3)}
\def\JHEP(#1,#2,#3){J.\ High Energy Physics \issue(#1,#2,#3)}
\def\JPG(#1,#2,#3){J.\ Phys.\ G \issue(#1,#2,#3)}
\def\MPL(#1,#2,#3){Mod.\ Phys.\ Lett.\ \issue(#1,#2,#3)}
\def\NP(#1,#2,#3){Nucl.\ Phys.\ \issue(#1,#2,#3)}
\def\NIM(#1,#2,#3){Nucl.\ Instrum.\ Meth.\ \issue(#1,#2,#3)}
\def\PL(#1,#2,#3){Phys.\ Lett.\ \issue(#1,#2,#3)}
\def\PRD(#1,#2,#3){Phys.\ Rev.\ D \issue(#1,#2,#3)}
\def\PRL(#1,#2,#3){Phys.\ Rev.\ Lett.\ \issue(#1,#2,#3)}
\def\SJNP(#1,#2,#3){Sov.\ J. Nucl.\ Phys.\ \issue(#1,#2,#3)}
\def\ZPC(#1,#2,#3){Zeit.\ Phys.\ C \issue(#1,#2,#3)}

\def\beq{\begin{equation}}
\def\eeq{\end{equation}}
\def\bea{\begin{eqnarray}}
\def\eea{\end{eqnarray}}
\def\to{\rightarrow}

\begin{document}


\preprint{
\vbox{
\hbox{IPMU11-0076, NPAC-11-10,SISSA-48/2011/EP} 
}}

\title{Electroweak Beautygenesis: \\ 
From $b \to s$ CP-violation to the  Cosmic Baryon Asymmetry }
\author{Tao Liu}
\affiliation{Department of Physics, University of California,
Santa Barbara, CA 93106, USA}
\author{Michael J. Ramsey-Musolf}
\affiliation{Department of Physics, University of Wisconsin, Madison, WI 53706, USA\\ and Kellogg Radiation Laboratory, California Institute of Technology, Pasadena, CA 91125}
 \author{Jing Shu}
\affiliation{Institute for the Physics and Mathematics of the Universe, University of Tokyo, Kashiwa, Chiba 277-8568, Japan}
\affiliation{International School for Advanced Studies (SISSA), Via Bonomea 265, I-34136 Trieste, Italy}

\begin{abstract}
We address the possibility that  CP-violation in $B_s-\bar B_s$ mixing may help explain the origin of the cosmic baryon asymmetry. We propose a new baryogenesis mechanism - \lq\lq Electroweak Beautygenesis" -- explicitly showing that these two CP-violating phenomena can be sourced by a common CP-phase. As an illustration,
we work in the Two-Higgs-Doublet model. Because the relevant CP-phase is flavor off-diagonal, this mechanism is less severely constrained
by null results of electric dipole moment searches than other scenarios. We show how measurements of flavor observables by the D0, CDF, and
LHCb collaborations test this scenario.

\end{abstract}

\maketitle

\noindent {\bf Introduction} 
The baryon asymmetry of the Universe (BAU) has been precisely
measured by the WMAP collaboration. Combining its five year results with those from other CMB
and large scale structure measurements gives
$\Omega_b h^2 = 0.02265\pm 0.00059$~\cite{Komatsu:2008hk} which is
in excellent agreement with the 95\% C.L. range $0.017-0.024$ obtained from 
Big Bang Nucleosynthesis~\cite{Nakamura:2010zzi}. The implied ratio of baryon density $n_B$  to entropy $s$  is
${{n_B}/ {s} } =( 8.82\pm 0.23)\times 10 ^{-11}$.

To generate the observed BAU, three Sakharov criteria~\cite{Sak}
must be satisfied in the early Universe: (1) baryon number violation; (2) C and
CP violation; (3) a departure from thermal equilibrium (or CPT
violation). These requirements are not unconquerable, though doing so requires physics beyond the Standard Model (SM) of particle physics. 
Indeed, there exist a number of possibilities, though none have been conclusively established.  One of the most popular --  standard thermal leptogenesis -- provides a theoretically attractive solution, yet it is generally difficult to test experimentally. It is, therefore, worth considering scenarios that may be more directly tested laboratory experiments. 

A particularly interesting and largely unexplored possibility involves CP-violation that enters the $B_s$ meson system. The relevant phases are generically
flavor off-diagonal, making them less susceptible to constraints from searches for permanent electric dipole moments (EDMs) that challenge other baryogenesis scenarios (for an illustration in the Minimal Supersymmetric Standard Model (MSSM), see, {\em e.g.} \cite{Li:2010ax}) . Moreover, recent measurements in B-factories and at the Tevatron  exhibit indications of CP-violation that differ by a few standard deviations from the SM predictions \cite{newbetas, dimuon}, though even more recent results from the LHCb collaboration favor smaller effects \cite{LHCbnewbetas}. From a theoretical perspective if the CP phase(s) encoding the CP-violation in $B_s$ system can successfully drive the generation of the BAU and can be probed experimentally,  our understanding of the BAU problem will be considerably advanced.

In this letter, we report on an initial effort addressing this question. We propose a new mechanism in the framework of Electroweak Baryogenesis (EWBG), explicitly showing that the CP-violating phenomena characterized by different energy scales ($B_s$ observables and BAU) can be sourced by a common CP-phase. As an illustration, we will work in a Two-Higgs-Doublet Model (2HDM). In this context, if a sufficiently strong, first-order electroweak phase transition (EWPT) occurs in the early Universe, the CP phase associated with the tree-level, Higgs-$b$-$s$ interaction at the phase boundary can induce CP-asymmetries that ultimately induce the BAU.  While EWBG in the 2HDM has been discussed extensively in the past~\cite{Fromme:2006cm},  including two recent studies using the 2HDM that have addressed the possible connection between the BAU and $B_s$ observables\cite{Tulin:2011wi} (see also \cite{Chao:2010mq}), we emphasize that the  mechanism discussed below is the only one thus far to explore the feasibility of baryogenesis directly driven by the $b \to s$ CP-violation. Given the novel features that are generically absent  elsewhere and the crucial role played by ``beauty'' quarks, we denote this mechanism ``Electroweak Beautygenesis'' (EWBTG). 

In what follows, we concentrate on the issue of CP-violation and do not treat the question of the first order EWPT in the 2HDM. Following Ref.~\cite{Tulin:2011wi}, we instead refer the reader to more general studies that may indicate its feasibility~\cite{Fromme:2006cm}. We note, however, that these analyses are typically gauge-dependent and therefore open to question. Rather than delve into these subtleties of perturbative treatments of the EWPT, we also refer the readers to a recent and more extensive discussion \cite{Patel:2011th}.


\noindent {\bf Two Higgs Doublet Model}
The Higgs sector in the general 2HDM is ($H_{u,d}$ are Higgs doublets with their SM gauge charges being $(0, 2, \pm 1/2)$)
\bea
\mathcal{L} =   \lambda^u_{ij} \bar{Q}^i (\epsilon H_d^\dagger) u^j_R -  \lambda^d_{ij} \bar{Q}^i H_d d^j_R  \nonumber \\ -  y_{ij}^u \bar{Q}^i H_u u^j_R  + y_{ij}^d \bar{Q}^i (\epsilon H_u^\dagger) d^j_R   + h.c. .
\label{101}
\eea 
In a supersymmetric embedding, the first term can arise at loop level~\cite{Dobrescu:2010mk}. 
For experimental relevance, we focus on the two flavor $b$-$s$ system, 
with its mass matrix 
\be
m_{ij} = y_{ij} v_u + \lambda_{ij} v_d= (y_{ij} \sin \beta + \lambda_{ij} \cos \beta) v ,
\ee 
where $v_{u,d}$ are vacuum expectation values (VEVs) of the neutral Higgs fields with $v = \sqrt{v_u^2 + v_d^2}$ and $\tan \beta = v_u / v_d$.
$v_{u,d}$ are functions of spacetime during the EWPT. Meanwhile, $H_{bs} = -\cos\beta H_u +\sin\beta H_d^\dagger$, a linear combination of Higgs mass eigenstates, 
will introduce flavor-changing neutral current effects at zero temperature.

Since we are investigating the feasibility that a common phase can source the BAU and account for the $B_s$ CP-violating observables, we will work in a simplified but sufficiently representative scenario, deferring a more comprehensive treatment to future work where the following scenario would arise in one region of parameter space. 
First, we take $\tan\beta = 1$ at zero temperature, emphasizing that  $\tan \beta$ is not a constant during an EWPT.
Second, we assume $y_{sb} = \lambda_{sb} = m_{sb} =0$.
In the limit of $y_{ss}, \lambda_{ss} \to 0$, there is one CP-violating phase after appropriate field redefinitions. Without loss of generality, we assume that $\lambda_{bs}$ is complex (with $\theta_{\lambda_{bs}}=  \textrm{Arg} ( \lambda_{bs})$) and $y_{bs}$, $y_{bb}$ and $\lambda_{bb}$ are real, and furthermore, assume $\lambda_{ii} = y_{ii}$ and $|\lambda_{bs}| = |y_{bs}|$.
The mass matrix is then
\bea 
\label{eq:mass}
\begin{pmatrix} \pm 2 \xi_{ss} & 0 \\ \xi_{bs} (\pm 1 + e^{i \theta_{\lambda_{bs}}} ) & \pm 2 \xi_{bb} \end{pmatrix}  v
  \, , 
\eea
here $\xi_{ij} = |\lambda_{ij}|/\sqrt{2} $ and the ``$\pm$'' signs are due to $y_{ss}$, $y_{bs}$ and $y_{bb} >$ or $<0$. 
Denoting $m_{bs}$ as $m_{bs} = \Delta \exp(i\theta)$, 
we have $\Delta = 2 \xi_{bs} |\cos(\theta_{\lambda_{bs}} /2)| v$, $\theta = \theta_{\lambda_{bs}}/2$ for $y_{bs} > 0$, and $\Delta = 2 \xi_{bs} | \sin(\theta_{\lambda_{bs}} /2)| v$, $\theta  = (\theta_{\lambda_{bs}}+ \pi )/2$ for $y_{bs} < 0$.

The mass matrix can be diagonalized by a unitary transformation $U_L^\dag M U_R = D$. 
In the small $m_{ss}$ limit, $U_L$ is trivial and $U_R$ is parametrized by a rotation angle $\alpha_R =  - \arctan (\Delta/m_{bb})$. The coupling of $H_{bs}$ and $b_L, s_R$ quarks in the mass eigenstate basis is given by
\bea
\zeta_{bs} = \xi_{bs}[\mp 1+ \exp(i \theta_{\lambda_{bs}})] \cos {\alpha_R}.
\eea
with 
$\textrm{Arg}(\zeta_{bs}) = \theta \pm \pi/2$ for $y_{bs} > 0$ and $< 0$, respectively. 
It is just the phase $\theta$ (or $\theta_{\lambda_{bs}}$) that both introduces CP-violation in $b\to s$ transitions (via $\zeta_{bs}$) and source the generation of baryon asymmetry (via  $m_{bs}$).

\noindent {\bf Electroweak Beautygenesis}
Production of the BAU during a first-order EWPT involves a dynamic generation of CP-violating charge asymmetries through particle interactions  in the wall of nucleated bubbles. 
Those charge asymmetries are converted, via left-handed fermions ($n_L$), into the baryon asymmetry through the electroweak sphaleron transitions.
We ignore the wall curvature in our analysis so all relevant functions depend on the variable $\bar{z} = z + v_{w} t$. Here $v_{w}$ is the wall velocity;  $\bar{z} < 0$ and $> 0$ correspond to the unbroken and broken phases, respectively; and the boundary extends over $0 < \bar{z} < L_{w}$. As pointed out in~\cite{Riotto:1998zb}, the transport properties of particles during the EWPT is most appropriately treated using non-equilibrium quantum field theory.
Working in its closed time path formulation (for pedagogical discussions, see~\cite{Lee:2004we}) and under the  \lq\lq VEV-insertion" approximation (see, {\em e.g.}, Refs.~\cite{Tulin:2011wi,Riotto:1998zb,Lee:2004we,Chung:2009qs}), we compute the CP-violating source induced by the Higgs mediated process $b_L\to s_R \to b_L$. It is given by
\begin{eqnarray}
S^\CPv_{b_L}  
&=& -S^\CPv_{s_R} =
\frac{N_c\Delta(\bar z)^2}{\pi^2}  \dot \theta (\bar z)  \int_0^\infty\frac{dk\,k^2} {\omega_{b_L}\omega_{s_R}}  \\
&& \times {\rm Im} \Big\{\frac{(\mathcal{E}_{b_L}^* \mathcal{E}_{s_R}-k^2)(n_F(\mathcal{E}_{s_R}) - n_F(\mathcal{E}_{b_L}^*))}{(\mathcal{E}_{s_R} 
- \mathcal{E}_{b_L}^*)^2} \nonumber \\ && + \frac{(\mathcal{E}_{b_L} \mathcal{E}_{s_R}+k^2)(n_F(\mathcal{E}_{s_R}) + n_F(\mathcal{E}_{b_L}))}{(\mathcal{E}_{s_R} + \mathcal{E}_{b_L})^2}\Big\}\ \ \ . \nonumber 
\label{101}
\end{eqnarray}
Here $n_F(x)$$=$$1/(\exp(x)+1)$ is the Fermi distribution; 
$\mathcal{E}_{b_L,s_R}$$=\omega_{b_L,s_R} - i \Gamma_{b_L,s_R}$ are complex poles of the spectral function
with $\omega_{b_L,s_R}^2 = k^2 + m_{b_L,s_R}^2$; and $m_{b_L,s_R}$ and $\Gamma_{b_L,s_R}$ are thermal parameters. This source  corresponds to the \lq\lq A"-type terms in Eq.~(58) of \cite{Lee:2004we}, after properly accounting for temperature-independent vacuum contributions that are removed via normal ordering \footnote{We thank C. Lee for clarification of this point.}. 
The quantity
$\dot \theta(\bar z)=d\theta(\bar z) /dt$ is given by
\begin{eqnarray}
\dot \theta(\bar z) = \frac{- 2f(\bar z)  }{\Delta(\bar z)^2} {\rm sign}(y_{bs}) \xi_{bs}^2(\infty) \sin \theta_{\lambda_{bs}}\label{103}
\end{eqnarray}
with $f(\bar z)  = (\dot v_u(\bar z) v_d (\bar z) - v_u(\bar z) \dot v_d(\bar z)) \sim v_wv^2 \delta\beta/L_w$ 
being a function describing the relative variation of the Higgs VEVs across the bubble wall. 
Although analyses performed in the MSSM~\cite{Moreno:1998bq} indicate $\delta \beta \sim 
\mathcal O(10^{-2})$, a systematic analysis is absent in the 2HDM. Here, we will simply adopt $\delta \beta = -0.05$ (if $y_{bs}$ is complex and $\lambda_{bs}$ is real, we need $\delta \beta > 0$ to keep the sign of $f(\bar z)$ unchanged). Note, $S^\CPv_{b_L}$ is non-zero only within the moving bubble wall, where $\dot \theta(\bar z)\not=0$.

In contrast to EWBG driven by flavor-diagonal sources, the transport of both the second and third family particles is sourced by CP-violating terms. 
We define the number densities $\{Q_{1,2,3},U,D,C,S,T,B,H=H_u^++H_u^0-H_d^--H_d^0\}$ which correspond, respectively,  
to left-chiral quarks of different families, right-chiral up, down, charm, strange, top and bottoms, and Higgs bosons. Since all light quarks (except $b_L$ and $s_R$) are mainly 
produced by strong sphaleron processes and all quarks have similar diffusion constants, baryon number conservation on time-scales shorter than the inverse electroweak sphaleron rate implies the approximate constraints $Q_1=Q_2=-2U=-2D=-2C=-2B$ and $S+T+Q_3=0$. The set of Boltzmann equations is 
\begin{eqnarray}
\partial^\mu {Q_3}_\mu &=&   \Gamma_{m_t} \left({\xi_T}-{\xi_{Q_3}}\right)  
+\Gamma_t\left({\xi_T}-{\xi_H}-{\xi_{Q_3}}\right) \nonumber \\
&& +2\Gamma_{ss}\left({\xi_T}-{2\xi_{Q_3}}+{\xi_S}+8\xi_B\right) + S^\CPv_{b_L}   \nonumber \\
\partial^\mu T_\mu  &=&- \Gamma_{m_t} \left({\xi_T}-{\xi_{Q_3}}\right)  
-\Gamma_t\left({\xi_T}-{\xi_H}-{\xi_{Q_3}}\right)   \nonumber  \\
&&-\Gamma_{ss}\left({\xi_T}-{2\xi_{Q_3}}+{\xi_S}+8\xi_B\right) \nonumber \\
\partial^\mu \delta_\mu& = & -S^\CPv_{b_L},    \ \ \ \  (\mbox{with} \ \ \delta = S-B) \nonumber 
\eea
\bea
\partial^\mu H_\mu &=&  \Gamma_t\left({\xi_T}-{\xi_H}-{\xi_{Q_3}}\right) - 2 \Gamma_h H  \label{105}     \ .
\end{eqnarray}
Here $\partial^\mu = v_w\frac{d}{d\bar z} - D_a \frac{d^2}{d\bar z^2}$ in the planar bubble wall approximation with $D_a$ being a diffusion constant, and $\xi_a=n_a/k_a$ with $n_a$ and $k_a$ being the number density and the statistical factor of particle ``a''.   
Apart from the CP-violating sources, the interactions in Eq.~(\ref{105}) include (i) inelastic top Yukawa ($\Gamma_t$) and strong sphaleron ($\Gamma_{ss}$) processes; 
(ii) top relaxation processes ($\Gamma_{m_t}$), while we neglect the other Yukawa interactions since $\Gamma D_q/v_w^2 < 1$; and (iii) Higgs relaxation processes ($\Gamma_h$) due to Higgs mass mixing, with typically $\Gamma_h < \Gamma_{m_t}$ in this scenario.  

Assuming $S^\CPv_{b_L}(\bar z <0)=0$, we solve the Boltzmann equations for the net left-handed fermion density $n_L=\sum_{i=1}^3Q_i$ analytically order-by-order in $1/\Gamma_{ss}$, with $\Gamma_{ss} = 16 \alpha_s^4 T$. The leading contribution arises at first order in this expansion. 
The baryon asymmetry $\rho_B$ is then produced in weak sphaleron process, described by~\cite{Cline:2000nw}
\begin{eqnarray}
\partial^\mu {\rho_B}_\mu = -\Theta(-\bar z) \Gamma_{ws} \left(\frac{15}{4} \rho_B +3 n_L\right) ,   \label{107}
\end{eqnarray}
where $\Gamma_{ws}=120\alpha_w^5 T$ is the weak sphaleron rate~\cite{Bodeker:1999gx}.  
In the broken phase this gives ($k_S=k_B$ is assumed)
\begin{eqnarray}
\rho_B =\frac{3 \Gamma_{\rm ws} }{v_w^2 }  
\int_{0}^{\infty}  \left[r \frac{ v_w^2}{\Gamma_{ss} \bar{D} }\left(1- \frac{D_q}{\bar{D}} \right)   \frac{\bar{S}(\bar z) }{ \kappa^+} e^{- \kappa^+  \bar z} \right] d\bar z \label{108}
\end{eqnarray}
with $r = - \frac{3}{2} \Big[ \frac{k_B (k_Q+2 k_T)}{k_H(9k_T+9k_Q+k_B)} \Big]$ and $\kappa^+ \simeq (\sqrt{v_w^2 + 4 \bar{\Gamma}  \bar{D}} + v_w)/2 \bar{D}$. Here $\bar D$, $\bar \Gamma$ and $\bar S$ are, respectively, the effective diffusion constant, decay rate and CP-violating source for  
the Higgs number density. $\bar D$ is defined in~\cite{Lee:2004we}, while 
\begin{eqnarray}
{\bar \Gamma}  &=& (9k_T+9k_Q+k_B ) (\Gamma_{m_t}+2\Gamma_h) /X
\nonumber \\
{\bar S}  &=& k_H (k_{Q_3}-7k_T+k_B)S_{b_L}^{\CPV}  /X  \\
X&=&9k_{Q_3}k_T + k_Bk_{Q_3} + 4k_T k_B +k_H(9k_T+9k_{Q_3}+k_B).  \nonumber \label{109}
\end{eqnarray}  
Note that while the weak sphaleron transitions are driven by the diffusion tail for $n_L$ that extends ahead of the advancing wall in the unbroken phase (${\bar z}< 0$), the solution in Eq.~(\ref{108}) contains an integral over the source in the broken phase that appears when matching the solutions to the Boltzmann equations at the phase boundary.

\begin{figure}[ht]
\setlength{\abovecaptionskip}{-5pt}
\setlength{\belowcaptionskip}{-5pt}
\begin{center}
    \includegraphics[width=7cm]{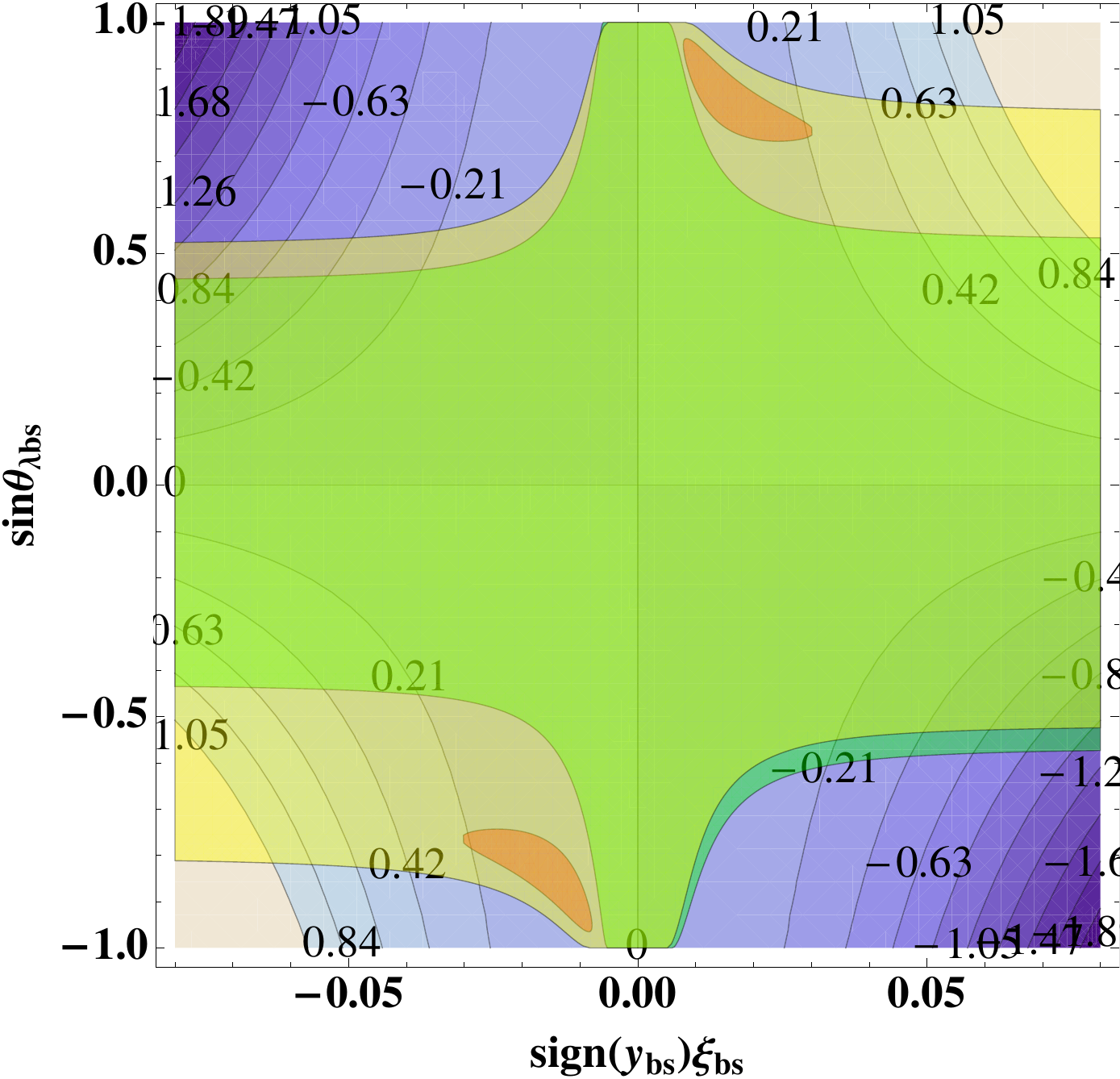} \\
\end{center}
  \caption{Contours of ${n_B} / {s}$ (in units of $10^{-10}$).
Orange, yellow and green contours indicate 95\% C. L.  Tevatron (with and without di-muon asymmetry) and LHCb constraints, respectively. We assume  $k_{Q_3}=2k_T=2k_B=6$~\cite{Huet:1995sh}, $k_H=4$, $v_w=0.4$~\cite{Megevand:2009gh}, $L_w = 2/T$, $D_q=6/T$ and $\Lambda_{bs}=1$ TeV.}  
  \label{fig_BAU}
\end{figure}

\noindent {\bf CP-violation in the $B_s-\bar B_s$ Mixing} 
Depending on the details of the scalar potential,  $H_{bs}$ may be approximately a mass eigenstate, which we assume for illustration. 
Tree-level exchange of $H_{\rm bs}$ with a VEV insertion 
leads to a $B_s-\bar B_s$ mixing operator in the basis of quark mass eigenstates ($\Lambda_{bs}$ is an effective new physics scale):
\begin{eqnarray}
&\frac{ \zeta_{bs}^2}{\Lambda_{bs}^2} (\bar{b}_L s_R)(\bar{b}_L s_R), \ \  {\rm with}  \ \  m_{H_{bs}}^2 \sim  v \Lambda_{bs}   \nonumber   
\label{operator}
\end{eqnarray}
The RG running of this operator involves a mixture of scalar operator $O_{SRR}^{bs}  \equiv (\bar{b} P_{R} s)( \bar{b} P_R s)$ and tensor operators $O_{TRR}^{bs}  \equiv (\bar{b} \sigma^{\mu \nu} P_{R} s)(\bar{b} \sigma_{\mu \nu} P_R s$) whose matrix elements are $\langle B_s | O_{SRR}^{bs} | \bar{B}_s \rangle$ $\approx$ $- 5 m_{B_s} f_{B_s}^2 B^{bs}_{SRR}/24$ and $\langle B_s | O_{TRR}^{bs} | \bar{B}_s \rangle$ $\approx$ $-  m_{B_s} f_{B_s}^2 B^{bs}_{TRR}/2$  \cite{Buras:2001ra}.
Assuming $m_{B_s}^2$ $\approx$ $(m_b + m_s)^2$ and $B^{bs}_{SRR}$ $\simeq$ ${B}^{bs}_{TRR}$ $=$ $B_{B_s}$, 
we obtain $
M^{s}_{12}  \equiv  \langle B_s |  \mathcal{H}  | \bar{B}_s \rangle 
$ $=$ $-{ \zeta_{bs}^2  } f_{B_s}^2 m_{B_s} B_{B_s} ({5 \eta_{SRR}/ 24 }+$ ${ \eta_{TRR}/2} )/\Lambda_{bs}^2$, 
with  $\eta_{SRR} \approx 1.87$, $\eta_{TRR} \approx -0.01$~\cite{Buras:2001ra}. 
  
\begin{table}[h!]
\setlength{\abovecaptionskip}{0pt}
\setlength{\belowcaptionskip}{-12pt}
\begin{ruledtabular}
\begin{tabular}{ccc}
$\beta_s^{\rm{SM}}$  &  $\beta_s^{\rm{Tev}}$ \cite{newbetas} & $f_{B_{s}} \sqrt{{B_{B_s}}}$  \\ \hline
  
 $0.019 \pm 0.001$ & $0.27 \pm {0.15}$  &  $(275 \pm 13)$ MeV \\ \hline\hline
$(\Delta m_{B_s})^{\rm SM}$ & $(\Delta m_{B_s})^{\rm Tev}$ \cite{Abulencia:2006ze}  & $\phi^{\rm SM}_s$ 
   \\ \hline
$19.30 \pm 2.2$ ps$^{-1}$ & $17.77 \pm 0.12$ ps$^{-1}$ & $(4.2 \pm 1.4) \times 10^{-3}$  \\ \hline

$(\Delta\Gamma_s)^{\rm SM}$ & $(\Delta \Gamma_s)^{\rm Tev}$ \cite{newbetas}   &  $A_{sl}^b$ \cite{dimuon} \\ \hline
$0.098 \pm 0.024 \ {\rm ps^{-1} }$ & $0.097 \pm {0.031}$ ps$^{-1}$  & $(- 7.40 \pm 1.93) \times 10^{-3}$
\\ \hline
$(\Delta m_{B_s})^{\rm LHCb}$ \cite{LHCbnewbetas} & $(\Delta \Gamma_s)^{\rm LHCb}$ \cite{LHCbnewbetas}  &  $\beta_s^{\rm{LHCb}}$ \cite{LHCbnewbetas}  \\ \hline
$17.725 \pm 0.049 \ {\rm ps^{-1} }$ & $0.123 \pm {0.030}$ ps$^{-1}$  & $-0.015 \pm 0.087$
\end{tabular}
\end{ruledtabular}
\caption{The theoretical input parameters \cite{Lenz:2006hd, Nakamura:2010zzi} and the experimental data from the Tevatron and LHCb. 
 } \label{tab:inputs}
\end{table}


Choosing $\Gamma_q$ to be real and parametrizing $M^s_{12}$ as \cite{Park:2010sg} 
$M_{12}^{s} \equiv (M_{12}^{s})^{\textrm{SM}}  \Delta_{s}$ with 
$ \Delta_s \equiv | \Delta_{s} | e^{i \phi_{s}^\Delta}$,
we have 
\bea
\label{eq:parametrization}
\Delta\Gamma_s = \Delta\Gamma_s^{\rm SM}\,
  \cos (\phi_s^{\rm SM} + \phi_s^\Delta ) \,,  && \Delta m_{s} = \Delta m_{s}^{\rm SM}\,
  \big| \Delta_{s} \big| \,, \nn\\
a^{s}_{\rm SL} =   \frac{ \Delta\Gamma_{s}^{\rm SM} }  { \Delta m_{s}^{\rm SM} } \frac{\sin (\phi_{s}^{\rm SM} + \phi_{s}^\Delta ) }{ | \Delta_{s} | } \,, && 2 \beta_s =  2\beta^{\rm SM}_s - \phi_s^\Delta  \, .
\eea
Here $\Delta m_{s}$ and $\Delta \Gamma_{s}$ are the mass and decay width difference between the heavy and light $B_s$ mass eigenstates, $a_{SL}^{s}$ is the charge asymmetry in semileptonic $B_{s}$ decays, and  $\beta_s$ measures the time-dependent CP asymmetries in the hadronic $B_s$ decay.

The theoretical inputs and experimental results are listed in the Table \ref{tab:inputs}. 
The decay constants and bag parameters are taken from Ref.~\cite{Laiho:2009eu}, 
while $A_{sl}^b$, and $\beta_s^{\rm{Exp}}$, $\Delta \Gamma^{\rm{Exp}}$ are obtained by combining the D$\cancel{\rm 0}$ and CDF measurements~\cite{newbetas, dimuon}. 
We perform a $\chi^2$ fit  to the four observables in Eq.(\ref{eq:parametrization}),
neglecting the correlation between $\Delta \Gamma_s$ and $\beta_s$  for simplicity.
Assuming $\Lambda_{bs}$ of 1 TeV, we scan over the remaining parameters, yielding the regions of 95\% C. L. from the Tevatron and the LHCb results.

\noindent {\bf Discussion}
The contours of constant ${n_B} / {s}$ in the ${\rm sign}(y_{bs})\xi_{bs}-\sin\theta_{\lambda_{bs}}$ plane are indicated in Fig.~\ref{fig_BAU}. We observe that the regions favored by the low-energy flavor studies at 95\% C.L. overlap with regions in which a sizable portion of the baryon asymmetry is generated. 
The LHCb results on the $B_s$ hadronic decay are more constraining on the parameter space than the Tevatron ones that do not include the dimuon asymmetry. Although tension exists between the LHCb- and Tevatron-favored regions, it appears feasible that a common CP-violating phase may be responsible for both generating a non-negligible portion of the BAU and accounting for observations in the $B_s$ system.  

A definitive statement awaits the resolution of both the experimental tensions as well as several theoretical issues, including the development of a VEV-resummed CP-violating source 
(for recent progress, see, {\em e.g.}, \cite{Konstandin:2005cd}), analysis of the full numerical solutions to Eqs. (\ref{105}), and completion of a gauge-invariant analysis of the EWPT in the 2HDM. Indeed, the results of this initial study are likely to indicate the maximum magnitude of the BAU that can be achieved in this scenario, given the generous assumptions we have made about various input parameters, including $\delta\beta$ and $v_w$ and the use of an analytic rather than numerical solution of the Boltzmann equations. 
Nevertheless, we expect that after future refinements are implemented, EWBTG may account for an interesting portion of the BAU in appropriate regions of parameter space. It thus appears that a more comprehensive analysis of this scenario is warranted.

Though the foregoing discussion relied on the illustrative case of a two-flavor system of the 2HDM with a single phase, generalization to
variants, including minimal flavor violation with flavor-blind phases ({\em e.g.}, see~\cite{Buras:2010mh}) or spontaneous CP-violation, would be straightforward. We leave the consideration of these possibilities, along with EWBTG in other models such as  the four-family SM ({\em e.g.}, see~\cite{Hou:2011de}), family non-universal $U(1)'$ model~\cite{Langacker:2000ju}, and supersymmetric models, {\em etc.} to future work. 

\noindent{\it  Acknowledgements} 
We thank V. Cirigliano, J. Cline, S. Gori, C. Lee, S. Tulin and C. Wagner for helpful discussions;  P. Draper for collaboration at the early stages of this project;  and the Shanghai Jiao Tong University where part of this work was carried out. We are grateful to S. Tulin for pointing out an error in an earlier version of this work and for a critical reading of this manuscript. 
The work is partially supported by U.S. Department of Energy contracts DE-FG02-91ER40618 (TL) and DE-FG02-08ER41531 (MJRM); the Wisconsin Alumni Research Foundation (MJRM); and the World Premier International Research Center Initiative (WPI initiative) MEXT, Japan (JS), Grant-in-Aid for scientific research (Young Scientists (B) 21740169) from Japan Society for Promotion of Science (JS) and the ERC Advanced Grant no. 267985 ``DaMESyFla'' (JS).


\end{document}